# Providing Secrecy with Lattice Codes

Xiang He    Aylin Yener
Wireless Communications and Networking Laboratory
Electrical Engineering Department
The Pennsylvania State University, University Park, PA 16802
xxh119@psu.edu    yener@ee.psu.edu

*Abstract*—Recent results have shown that lattice codes can be used to construct good channel codes, source codes and physical layer network codes for Gaussian channels. On the other hand, for Gaussian channels with secrecy constraints, efforts to date rely on random codes. In this work, we provide a tool to bridge these two areas so that the secrecy rate can be computed when lattice codes are used. In particular, we address the problem of bounding equivocation rates under nonlinear modulus operation that is present in lattice encoders/decoders. The technique is then demonstrated in two Gaussian channel examples: (1) a Gaussian wiretap channel with a cooperative jammer, and (2) a multi-hop line network from a source to a destination with untrusted intermediate relay nodes from whom the information needs to be kept secret. In both cases, lattice codes are used to facilitate cooperative jamming. In the second case, interestingly, we demonstrate that a non-vanishing positive secrecy rate is achievable regardless of the number of hops.

## I. INTRODUCTION

Information theoretic secrecy was first proposed by Shannon in [1]. In this classical model, Bob wants to send a message to Alice, which needs to be kept secret from Eve. Shannon's notion of secrecy requires the average rate of information leaked to Eve to be zero, with no assumption made on the computational power of Eve. Wyner, in [2], pointed out that, more often than not, the eavesdropper (Eve) has a noisy copy of the signal transmitted from the source, and building a useful secure communication system per Shannon's notion is possible [2]. Csiszar and Korner [3] extended this to a more general channel model.

Numerous channel models have since been studied under Shannon's framework. The maximum reliable transmission rate with secrecy is identified for several cases including the Gaussian wiretap channel [4] and the MIMO wiretap channel [5], [6], [7]. Sum secrecy capacity for a degraded Gaussian multiple access wiretap channel is given in [8]. For other channels, upper bounds, lower bounds and some asymptotic results on the secrecy capacity exist. For the achievability part, Shannon's random coding argument proves to be effective in majority of these works.

On the other hand, it is known that the random coding argument may be insufficient to prove capacity theorems for certain channels [9]. Instead, structured codes like lattice codes are used. Using structured codes has two benefits. First, it is relatively easy to analyze large networks under these codes. For example, in [10], [11], the lattice code allows the relaying scheme to be equivalent to a modulus sum operation, making it easy to trace the signal over a multi-hop relay network. Secondly, the structured nature of these codes makes it possible to align unwanted interference, for example, for the interference channel with more than two users [12], [13], and the two way relay channel [10], [11].

A natural question is therefore whether structured codes are useful for secure communication as well. In particular, in this work, we are interested in answering two questions:
1) How do we bound the secrecy capacity when structured codes are used?
2) Are there models where structured codes prove to be useful in providing secrecy?

Relevant references in this line of thinking includes [14] and [15]. Reference [14] considers a binary additive two-way wiretap channel where one terminal uses binary jamming signals. Reference [15] examines a wiretap channel where the eavesdropping channel is a modulus-$\Lambda$ channel. Under the proposed signaling scheme therein, the source uses a lattice code to convey the secret message, and, the destination jams the eavesdropper with a lattice code. The eavesdropper sees the sum of these two codes, both taking value in a finite group, where the sum is carried under the addition defined over the group. It is known that if the jamming signal is sampled from a uniform distribution over the group, then the sum is independent from the message.

While these are encouraging steps in showing the impact of structured jamming signals, as commented in [15], using this technique in Gaussian channels is a non-trivial step. In the Gaussian channel, also, the eavesdropper receives the sum of the signal from the source and the jamming signal. However, the addition is over real numbers rather than over a finite group. The property of modulus sum is therefore lost and it is difficult to measure how much information is leaked to the eavesdropper.

Most lattice codes for power constrained transmission have a similar structure to the one used in [15]. First, a lattice is constructed, which should be a good channel code under the noise/interference. Then, to meet the power constraint, the lattice, or its shifted version, is intersected with a bounded set, called the shaping set, to create a set of lattice points with finite average power. The lattice is shifted to make sure sufficiently many lattice points fall into the shaping set to maintain the codebook size and hence the coding rate [16]. The decoder at the destination is called a lattice decoder if it is only asked to find the most likely lattice point under the received signals, and is not aware of shaping set. Because of the structured nature of the lattice, a lattice decoder has lower complexity compared to the maximum likelihood decoder where the knowledge of shaping set is used. Also, under the lattice decoder, the introduction of shaping set does not pose any additional difficulty to the analysis of decoding performance. Commonly used shaping sets include the sphere [12] and the fundamental region of a lattice [17].

A key observation is that, from the viewpoint of an eavesdropper, the shaping set actually provides useful information, since it reduces the set of lattice points the eavesdropper needs to consider. The main aim of this work, therefore, is to find a shaping set and lattice code construction under which the information leaked to the eavesdropper can be bounded. This shaping set, as we shall see, turns out to be the fundamental region of a "coarse" lattice in a nested lattice structure. Under this construction, we show that at most 1 bit is leaked to the eavesdropper per channel use. This enables us to lower bound the secrecy rate using a technique similar to the genie bound from [18].

To demonstrate the utility of our approach, we then apply our technique to two channel models: a Gaussian wiretap channel with a cooperative jammer, and a multi-hop line network, where a source can communicate a destination only through a chain of untrusted relays. In the second case, we demonstrate that a non-vanishing positive secrecy rate is achievable *regardless of the number of hops*.

The following notation is used throughout this work: We use $H$ to denote the entropy. $\varepsilon_k$ is used to denote any variable that goes to 0 when $n$ goes to $\infty$. We define $C(x) = \frac{1}{2}\log_2(1+x)$. $\lfloor a \rfloor$ denotes the largest integer less than or equal to $a$.

## II. THE REPRESENTATION THEOREM

In this section, we present a result about lattice codes which will be useful in the sequel.

Let $\Lambda$ denote a lattice in $\mathcal{R}^N$ [17], i.e., a set of points which is a group closed under real vector addition. The modulus operation $x \mod \Lambda$ is defined as $x \mod \Lambda = x - \arg\min_{y \in \Lambda} d(x,y)$, where $d(x,y)$ is the Euclidean distance between $x$ and $y$. The fundamental region of a lattice $\mathcal{V}$ is defined as the set $\{x : x \mod \Lambda = 0\}$. It is possible that there are more than one lattice points that have the same minimal distance to $x$. Breaking a tie like this is done by properly assign the boundary of $\mathcal{V}$ [17].

Let $t_A$ and $t_B$ be two numbers taken from $\mathcal{V}$. For any set $A$, define $2A$ as $2A = \{2x : x \in A\}$. Then we have:

$$\{t_A + t_B : t_A, t_B \in \mathcal{V}\} = 2\mathcal{V} \quad (1)$$

Define $A_x$ as $A_x = \{t_A + t_B + x, t_A, t_B \in \mathcal{V}\}$. Then from (1), we have $A_x = x + 2\mathcal{V}$. With this preparation, we are ready to prove the following *representation theorem*:

*Theorem 1:* There exists a random integer $T$, such that $1 \leq T \leq 2^N$, and $t_A + t_B$ is uniquely determined by $\{T, t_A + t_B \mod \Lambda\}$.

*Proof:* By definition of the modulus $\Lambda$ operation, we have

$$t_A + t_B \mod \Lambda = t_A + t_B + x, \quad x \in \Lambda \quad (2)$$

The theorem is equivalent to finding the number of possible $x$ meeting equation (2) for a given $t_A + t_B \mod \Lambda$.

To do that, we need to know a little more about the structure of lattice $\Lambda$. Every point in a lattice, by definition, can be represented in the following form [19]: $x = \sum_{i=1}^{N} a_i v_i$, $v_i \in \mathcal{R}^N, a_i \in Z$. $\{a_i\}$ is said to be the coordinates of the lattice point $x$ under the basis $\{v_i\}$.

Based on this representation, we can define the following relationship: Consider two points $x, y \in \Lambda$, with coordinates $\{a_i\}$ and $\{b_i\}$ respectively. Then we say $x \sim y$ if $a_i = b_i \mod 2$, $i = 1...N$. It is easy to see the relationship $\sim$ is an equivalence relationship. Therefore, it defines a partition over $\Lambda$.

1) Depending on the values of $a_i - b_i \mod 2$, there are $2^N$ sets in this partition.
2) The sub-lattice $2\Lambda$ is one set in the partition, whose members have even coordinates. The remaining $2^N - 1$ sets are its cosets.

Let $C_i$ denote any one of these cosets or $2\Lambda$. Then $C_i$ can expressed as $C_i = 2\Lambda + y_i$, $y_i \in \Lambda$. It is easy to verify that $A_x = x + 2\mathcal{V}$, $x \in C_i$ is a partition of $2\mathcal{R}^N + y_i$, which equals $\mathcal{R}^N$.

We proceed to use the two partitions derived above: Since $C_i, i = 1...2^N$ is a partition of $\Lambda$, (2) can be solved by considering the following $2^N$ equations:

$$t_A + t_B \mod \Lambda = t_A + t_B + x, \quad x \in C_i \quad (3)$$

From (1), this means $t_A + t_B \mod \Lambda \in x + 2\mathcal{V}$ for some $x \in C_i$. Since $x + 2\mathcal{V}, x \in C_i$ is a partition of $\mathcal{R}^N$, there is at most one $x \in C_i$ that meets this requirement. This implies for a given $t_A + t_B \mod \Lambda$, and a given coset $C_i$, (3) only has one solution for $x$. Since there are $2^N$ such equations, (2) has at most $2^N$ solutions. Hence each $t_A + t_B \mod \Lambda$ corresponds to at most $2^N$ points of $t_A + t_B$. ∎

*Remark 1:* Theorem 1 implies that modulus operation looses at most one bit per dimension of information if $t_A, t_B \in \mathcal{V}$.

The following crypto lemma is useful and is provided here for completeness.

*Lemma 1:* [15] Let $t_A, t_B$ be two independent random variables distributed over the a compact abelian group, $t_B$ has a uniform distribution, then $t_A + t_B$ is independent from $t_A$. Here $+$ is the addition over the group.

In the remainder of the paper, $(\Lambda, \Lambda_1)$ denotes a nested lattice structure where $\Lambda_1$ is the coarse lattice. Let $\mathcal{V}$ and $\mathcal{V}_1$ be their respective fundamental regions. We shall use $a \oplus b$, short for $a + b \mod \Lambda_1$. Then from Lemma 1, we have the following corollary:

*Corollary 1:* Let $t_A \in \Lambda \cap \mathcal{V}_1$. $t_B \in \Lambda \cap \mathcal{V}_1$ and $t_B$ is uniformly distributed over $\Lambda \cap \mathcal{V}_1$. Let $t_S = t_A \oplus t_B$. Then $t_S$ is independent from $t_A$.

## III. WIRETAP CHANNEL WITH A COOPERATIVE JAMMER

In this section, we demonstrate the use of lattice codes for secrecy in the simple model depicted in Figure 1. Nodes $S, D, E$ form a wiretap channel where $S$ is the source node, $D$ is the destination node, $E$ is the eavesdropper. Let the average power constraint of node $S$ be $P$. Now suppose that there is another transmitter $CJ$ in the system, also with power constraint $P$, as shown in Figure 1. We assume that the interference caused by $CJ$ to node $D$ is either too weak or too strong that it can be ignored or removed, and consequently there is no link between $CJ$ and $D$. In this model, node $CJ$ may *choose* to help $S$ by transmitting a jamming signal to confuse the eavesdropper $E$. Below, we derive the secrecy rate for this case when the jamming signal is chosen from a lattice codebook.

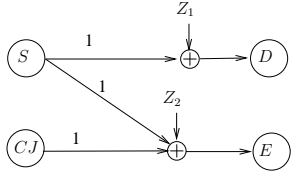

Fig. 1. Wiretap Channel with a Cooperative Jammer, CJ

### A. Gaussian Noise

We first consider the case when $Z_1$ and $Z_2$ are independent Gaussian random variables with zero mean and unit variance. In this case, we have the following theorem:

*Theorem 2:* A secrecy rate of $[C(P)-1]^+$ is achievable.

*Proof:* The codebook is constructed as follows: Let $(\Lambda, \Lambda_1)$ be a properly designed nested lattice structure in $\mathcal{R}^N$ as described in [17]. The codebook is all the lattice points within the set $\Lambda \cap \mathcal{V}_1$.

Let $t_A^N$ be the lattice point transmitted by node $S$. Let $d_A^N$ be the dithering noise uniformly distributed over $\mathcal{V}_1$. The transmitted signal is given by $t_A^N \oplus d_A^N$. The receiver receives the above signal corrupted by Gaussian noise and tries to decode $t_A^N$. Let the decoding result be $\hat{t}_A^N$. Then as shown in [17, Theorem 5], there exists a sequence of properly designed $(\Lambda, \Lambda_1)$ with increasing dimension, such that

$$\lim_{N\to\infty} \frac{1}{N} \log_2 |\Lambda \cap \mathcal{V}_1| < C(P) \quad (4)$$

$$C(P) = \frac{1}{2} \log_2(1+P) \quad (5)$$

and $\lim_{N\to\infty} \Pr(t_A^N \neq \hat{t}_A^N) = 0$.

The cooperative jammer $CJ$ uses the same codebook as node $S$. Let the lattice point transmitted by $CJ$ be $t_B^N$ and the dithering noise be $d_B^N$. The transmitted signal is given by $t_B^N \oplus d_B^N$. As in [17], we assume that $d_A^N$ is known by node $S$, the legitimate receiver node $D$ and the eavesdropper node $E$. $d_B^N$ is known by node $S$, and the eavesdropper node $E$. Hence, there is no common randomness between the legitimate communicating pairs that is not known by the eavesdropper.

Then the signal received by the eavesdropper can be represented as $t_A^N \oplus d_A^N + t_B^N \oplus d_B^N + Z_2^N$, where $Z_2^N$ is the Gaussian channel noise over $N$ channel uses. Then we have

$$H(t_A^N | t_A^N \oplus d_A^N + t_B^N \oplus d_B^N + Z_2^N, d_A^N, d_B^N) \quad (6)$$

$$\geq H(t_A^N | t_A^N \oplus d_A^N + t_B^N \oplus d_B^N + Z_2^N, d_A^N, d_B^N, Z_2^N) \quad (7)$$
$$= H(t_A^N | t_A^N \oplus d_A^N + t_B^N \oplus d_B^N, d_A^N, d_B^N) \quad (8)$$
$$= H(t_A^N | t_A^N \oplus d_A^N \oplus t_B^N \oplus d_B^N, d_A^N, d_B^N, T) \quad (9)$$
$$= H(t_A^N | t_A^N \oplus t_B^N, d_A^N, d_B^N, T) \quad (10)$$
$$= H(t_A^N | t_A^N \oplus t_B^N, T) \quad (11)$$
$$= H\left(T | t_A^N \oplus t_B^N, t_A^N\right) + H\left(t_A^N | t_A^N \oplus t_B^N\right) - H\left(T | t_A^N \oplus t_B^N\right) \quad (12)$$
$$\geq H\left(t_A^N | t_A^N \oplus t_B^N\right) - H\left(T | t_A^N \oplus t_B^N\right) \quad (13)$$
$$= H\left(t_A^N\right) - H\left(T | t_A^N \oplus t_B^N\right) \quad (14)$$
$$\geq H\left(t_A^N\right) - H\left(T\right) \quad (15)$$

In (9), we introduce the $N$ bit information $T$ that will help to recover $t_A^N \oplus d_A^N + t_B^N \oplus d_B^N$ from $t_A^N \oplus d_A^N \oplus t_B^N \oplus d_B^N$. In (14), we use the fact that $t_A^N$ is independent from $t_A^N \oplus t_B^N$ based on Corollary 1.

Let $c = \frac{1}{N} I\left(t_A^N; t_A^N \oplus d_A^N + t_B^N \oplus d_B^N + Z_2^N, d_A^N, d_B^N\right)$. Then from (15), since $H(T) \leq N$, we have $c \leq 1$. Therefore, if the message is mapped one-to-one to $t_A^N$, then an equivocation rate of at least $C(P) - 1$ is achievable under a transmission rate of $C(P)$ bits per channel use.

We note that to obtain perfect secrecy, some additional effort is required. First, we define a block of channel uses as the $N$ channel uses required to transmit a $N$ dimensional lattice point. A perfect secrecy rate of $C(P) - 1$ can then be achieved by coding across multiple blocks: A codeword in this case is composed of $Q$ components, each component is an $N$ dimensional lattice point sampled from a uniform distribution over $\mathcal{V}_1 \cap \Lambda$ in an i.i.d. fashion. The resulting codebook $\mathcal{C}$ contains $2^{\lfloor NQR \rfloor}$ codewords with $R < C(P)$. Like wiretap codes, the codebook is then randomly binned into several bins, where each bin contains $2^{\lfloor NQc \rfloor}$ codewords. The secret message $W$ is mapped to the bins. The actual transmitted codeword is chosen from that bin according to a uniform distribution.

Let $Y_e^{NQ}$ denote the signals available to the eavesdropper: $Y_e^{NQ} = \{t_A^{NQ} \oplus d_A^{NQ} + t_B^{NQ} \oplus d_B^{NQ} + Z^{NQ}, d_A^{NQ}, d_B^{NQ}\}$. Then we have

$$H(W|Y_e^{NQ}, \mathcal{C})$$
$$= H(W|t_A^{NQ}, Y_e^{NQ}, \mathcal{C}) + H(t_A^{NQ}|Y_e^{NQ}, \mathcal{C}) - H(t_A^{NQ}|W, Y_e^{NQ}, \mathcal{C}) \quad (16)$$
$$\geq H(t_A^{NQ}|Y_e^{NQ}, \mathcal{C}) - NQ\varepsilon \quad (17)$$
$$= H(t_A^{NQ}|Y_e^{NQ}, \mathcal{C}) - H(t_A^{NQ}|\mathcal{C}) + H(t_A^{NQ}|\mathcal{C}) - NQ\varepsilon \quad (18)$$
$$= H(t_A^{NQ}|\mathcal{C}) - I(t_A^{NQ}; Y_e^{NQ}|\mathcal{C}) - NQ\varepsilon \quad (19)$$
$$\geq H\left(t_A^{NQ}|\mathcal{C}\right) - \sum_{q=1}^{Q} I\left(t_A^N; Y_e^N|\mathcal{C}\right) - NQ\varepsilon \quad (20)$$
$$= H\left(t_A^{NQ}|\mathcal{C}\right) - QNc - NQ\varepsilon = QN(R-c) - NQ\varepsilon \quad (21)$$

In (17), we use Fano's inequality to bound the last term in (16). This is because the size of each bin is kept small enough such that given $W$, the eavesdropper can determine $t_A^{NQ}$ from its received signal $Y_e^{NQ}$. Using the standard random coding argument and (21), it can then be shown a secrecy rate of $C(P) - c$ is achievable. Since $c < 1$, this means a secrecy rate of at least $C(P) - 1$ bits per channel use is achievable. ∎

*Remark 2:* It is interesting to compare the secrecy rate obtained here with that obtained by cooperative jamming with Gaussian noise [20]. The latter is given by $C(P) - C(\frac{P}{P+1})$. $\lim_{P \to \infty} C(\frac{P}{P+1}) = 0.5$. Therefore there is at most 0.5 bit per channel use of loss in secrecy rate at high SNR by using a structured code book as the jamming signal.

### B. Non-Gaussian Noise

The performance analysis in [17] requires Gaussian noise. This is not always the case, for example, in the presence of interference, which is not necessarily Gaussian. For non-Gaussian noise, in principle, the analysis in [16] can be used instead. On the other hand, in [16], a sphere is used as the shaping set, making it difficult to computing the equivocation rate via Theorem 1. We show below, if the code rate $R$ has the form $\log_2 t, t \in \mathbf{Z}^+$, then a scaled lattice $t\Lambda$ of the fine lattice $\Lambda$ can be used for shaping instead.

*Theorem 3:* If $Z_1, Z_2$ are i.i.d. continuous random variables with differential entropy $h(E)$, such that $2^{2h(E)} = 2\pi e$, then a secrecy rate of $[\log_2 \lfloor \sqrt{P} \rfloor - 1]^+$ is achievable.

*Proof:* We need to show that there exists a fine lattice $\Lambda$ that has a good decoding performance [16, Theorem 6], and $\Lambda$ is close to a sphere in the sense that

$$\lim_{N \to \infty} h(S) = \frac{1}{2} \log_2(2\pi e P') \quad (22)$$

where $h(S) = \frac{1}{N} \log_2 |\mathcal{V}|$, $|\mathcal{V}|$ is the volume of the fundamental region of $\Lambda$, and $P' = \frac{1}{N|\mathcal{V}|} \int_{x \in \mathcal{V}} \|x\|^2 dx$. It is shown in [21] that when a lattice is sampled from the lattice ensemble defined therein, it is close to a sphere in the sense of (22). The lattice ensemble is generally called construction A [16], whose generation matrices are all matrix of size $K \times N$ over finite group $\mathrm{GF}(q)$, with $q$ being a prime. The lattice sampled from the ensemble is "good" in probability when $q, N \to \infty$ and $K$ grows faster than $\log^2 N$ [21, (25)-(28)]. Note that this property of "goodness" is invariant under scaling. Therefore, we can scale the lattice so that the volume of its fundamental region remains fixed when its dimension $N \to \infty$. This gives us a sequence of lattice ensembles that meet the condition of [13, Lemma 1]: (1) $N \to \infty$ (2) $q \to \infty$. (3) Each lattice ensemble of a given dimension is balanced [16]. This means when $N \to \infty$, at least 3/4 of the lattice ensemble is good for channel coding [13, Lemma 1]. The lattice decoder will have a positive decoding error exponent as long as $|\mathcal{V}| > 2^{Nh(E)}$. Combined, this means there must exist a lattice $\Lambda^*$ that is close to a sphere and is a good channel code at the same time. Hence we have $\frac{1}{N} \log_2 |\mathcal{V}| \to \frac{1}{2} \log_2(2\pi e P')$ as $N \to \infty$. Since we assume $h(E) = \frac{1}{2} \log_2(2\pi e)$ and require $|\mathcal{V}| > 2^{Nh(E)}$, this means as long as $P' > 1$, the decoding error will decrease exponentially when $N \to \infty$.

Now pick the shaping set to be the fundamental region of $t\Lambda^*, t \in \mathbf{Z}^+$. Then the code rate $R = \log_2(t)$ [17]. With the dithering and modulus operation from [17], the average power of the transmitted signal per dimension is $t^2 P'$. Note that the modulus operation at the destination, required in order to remove the dithering noise, may distort the additive channel noise. However, the decoding error event, defined as the noise pushing a lattice codeword into the set of typical noise sequence centered on a different lattice point [16], remains identical. Therefore, the decoding error exponent is the same. Hence we have $P' > 1$ and $t^2 P' < P$. The largest possible $t$ is $\lfloor \sqrt{P} \rfloor$, with the rate being $\log_2(\lfloor \sqrt{P} \rfloor)$. With similar arguments as in Theorem 2, we conclude that a secrecy rate of $[\log_2(\lfloor \sqrt{P} \rfloor) - 1]^+$ is achievable. ∎

## IV. MULTI-HOP LINE NETWORK WITH UNTRUSTED RELAYS

### A. System Model

In this section, we examine a more complicated communication scenario, as shown in Figure 2. The source has to communicate over $K - 1$ hops ($K \geq 3$) to reach the destination. Yet the intermediate relaying nodes are untrusted and need to be prevented from decoding the source information. Under this model, we will show that, using Theorem 1, with lattice codes for source transmission and jamming signals and an appropriate transmission schedule, an end-to-end secrecy rate that is independent of the number of untrusted relay nodes is achievable. We assume nodes can not receive and transmit signals simultaneously. We assume that each node can only communicate to its two neighbors, one on each side. Let $Y_i$ and $X_i$ be the received and transmitted signal of the $i$th node respectively. Then they are related as $Y_i = X_{i-1} + X_{i+1} + Z_i$, where $Z_i$ are zero mean Gaussian random variables with unit variance, and are independent from each other. Each node has the same average power constraint: $\frac{1}{n} \sum_{k=1}^{n} E[X_i(k)^2] \leq \bar{P}$ where $n$ is the total number of channel uses. The channel gains are normalized for simplicity.

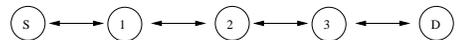

Fig. 2. A Line Network with 3 Un-trusted Relays

We consider the case where there is an eavesdropper residing at each relay node and these eavesdroppers are not cooperating. This also addresses the scenario where there is one eavesdropper, but the eavesdropper may appear at any one relay node that is unknown a priori. In either case, we need secrecy from all relays and the secrecy constraints for the $K$ relay nodes are expressed as $\lim_{n \to \infty} \frac{1}{n} H(W | Y_i^n) = \lim_{n \to \infty} \frac{1}{n} H(W), i = 1...K$.

### B. Signaling Scheme

Because all nodes are half duplex, a schedule is necessary to control when a node should talk. The node schedule is best represented by the acyclic directional graph as shown in Figure 3. The columns in Figure 3 indicate the nodes and the rows in Figure 3 indicate the phases. The length of a phase is the number of channel uses required to transmit a lattice point, which equals the dimension of the lattice. A node in a row has an outgoing edge if it transmits during a phase. The node in that row has an incoming edge if it can hear signals during the previous phase. It is understood, though not shown in the figure, that the signal received by the node is a superposition of the signals over all incoming edges corrupted by the additive Gaussian noise.

A number of consecutive phases is called one block, as shown in Figure 3. The boundary of a block is shown by the dotted line in Figure 3. The data transmission is carried over $M$ blocks.

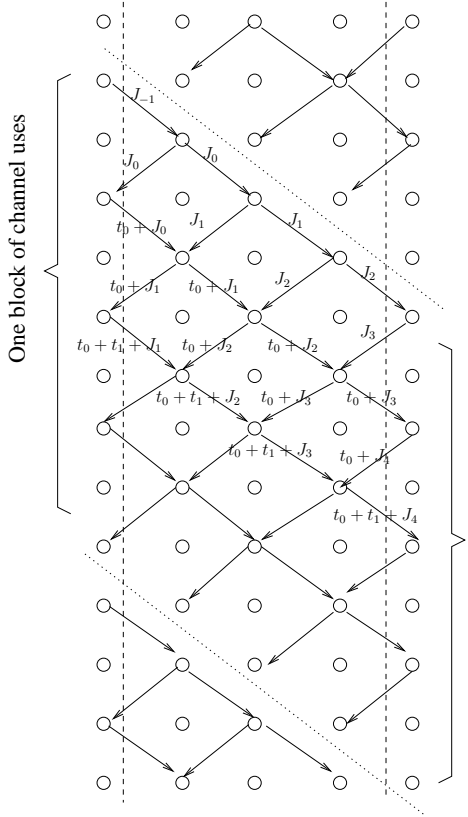

Fig. 3. One Block of Channel Uses

Again the nested lattice code $(\Lambda, \Lambda_1)$ from [10] is used within each block. The codebook is constructed in the same fashion as in Section III.

*1) The Source Node:* The input to the channel by the source has the form $t^N \oplus J^N \oplus d^N$. Here $d^N$ is the dithering noise which is uniformly distributed over $\mathcal{V}_1$. $t^N$ and $J^N$ are determined as follows: If it is the first time the source node transmits during this block, $t^N$ is the origin. $J^N$ is picked from the lattice points in $\Lambda \cap \mathcal{V}_1$ under a uniform distribution. Otherwise, $t^N$ is picked by the encoder. $J^N$ is the lattice point decoded from the jamming signal the source received during the previous phase. This design is not essential but it brings some uniformness in the form of received signals and simplifies explanation.

*2) The Relay Node:* As this signal propagates toward the destination, each relay node, when it is its turn, sends a jamming signal in the form of $t_k^N + d_k^N \mod \Lambda, k = 2...K-1$, where $K$ is the number of nodes. Subscript $k$ denotes the node index which transmit this signal. If this is the first time the relay transmits during this block, then $t_k^N$ is drawn from a uniform distribution over $\Lambda \cap \mathcal{V}_1$, and all previous received signals are ignored. Otherwise, $t_k^N$ is computed from the signal it received during the previous phase. This will be clarified in the sequel. $d_k^N$ again is the dithering noise uniformly distributed over $\mathcal{V}_1$.

The signal received by the relay within a block can be categorized into the following three cases. Let $z^N$ denote the Gaussian channel noise.

1) If this is the first time the relay receives signals during this block, then it has the form $(t_A^N \oplus d_A^N) + z^N$. It only contains interference from its left neighbor.
2) If this is the last time the relay receives signals during this block, then it has the form $(t_B^N \oplus d_B^N) + z^N$. It only contains interference from its right neighbor.
3) Otherwise it has the form $y_k^N = (t_A^N \oplus d_A^N) + (t_B^N \oplus d_B^N) + z^N$.

Here $t_A^N$, $t_B^N$ are lattice points, and $d_A^N$, $d_B^N$ are dithering noises. Following reference [10], if the lattice is properly designed and the cardinality of the set $\Lambda \cap \mathcal{V}_1$ is properly chosen, then for case (3), the relay, with the knowledge of $d_A^N, d_B^N$, will be able to decode $t_A^N \oplus t_B^N$. For case (1) and (2), the relay will be able to decode $t_A^N$ and $t_B^N$ respectively. Otherwise, we say that a decoding error has occurred at the relay node.

The transmitted signal at the relay node is then computed as follows:

$$x^N = t_A^N \oplus t_B^N \oplus (-x'^N) \oplus d_C^N \quad (23)$$

Here $x'^N$ is the lattice point contained in the jamming signal transmitted by this relay node during the previous phase. $-$ is the inverse operation defined over the group $\mathcal{V}_1 \cap \Lambda$. $t_A^N \oplus t_B^N$ are decoded from the signal it received during the previous phase.

In Figure 3, we labeled the lattice points transmitted over some edges. For clarity we omitted the superscript $^N$. The $+$ signs in the figure are all modulus operations. The reason why we have $(-x'^N)$ in (23) is now apparent: it leads to a simple expression for the signal as it propagates from the relay to the destination.

*3) The Destination:* As shown in Figure 3, the destination behaves identically to a relay node when it computes its jamming signal.

It is also clear from Figure 3 that the destination will be able to decode the data from the source. This is because the lattice point contained in the signal received by the destination has the form $t^N \oplus J^N$, where $t^N$ is the lattice point determined by the transmitted data, and $J^N$ is the lattice point in the jamming signal known by the destination.

### C. A Lower Bound to the Secrecy Rate

Suppose the source transmits $Q+1$ times within a block. Then each relay node receives $Q+2$ batches of signals within the block. An example with $Q = 2$ is shown in Figure 3. Given the inputs from the source of the current block, the signals received by the relay node are independent from the signals it received during any other block. Therefore, if a block of channel uses is viewed as one meta-channel use, with the source input as the channel input and the signal received by the relay as the channel output, then the effective channel is memoryless. Each relay node has the

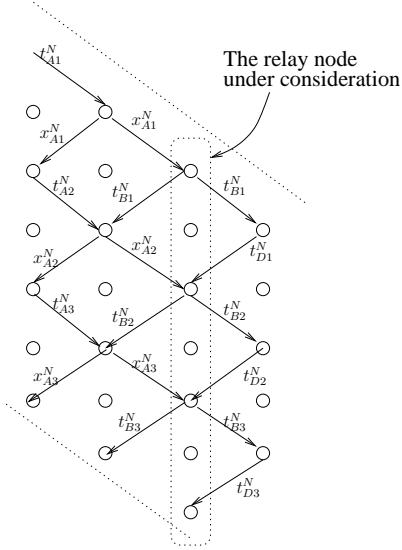

Fig. 4. Notations for Lattice Points contained in Signals, $Q = 2$

following side information regarding the source inputs within one block:

1) $Q + 2$ batches of received signals.
2) All the dithering noises $\{d_i\}$.
3) Signals transmitted from the relay node during this block. Note that only the first batch of signals it transmitted may provide information because all subsequent transmitted signals are computed from received signals and dithering noises.

Let $W$ be the secret message transmitted over $M$ blocks. Following the notation in Figure 4, the equivocation with respect to the relay node is given by:

$$H_2 = \frac{1}{NM} H(W | (x_{A1}^{NM} \oplus d_{\alpha 1}^{NM}) + z_1^{NM}, d_{\alpha 1}^{NM}$$
$$(x_{Ai}^{NM} \oplus d_{\alpha i}^{NM}) + (t_{D(i-1)}^{NM} \oplus d_{\beta(i-1)}^{NM}) + z_i^{NM},$$
$$d_{\alpha i}^{NM}, d_{\beta(i-1)}^{NM}, i = 2...Q+1$$
$$(t_{D(Q+1)}^{NM} \oplus d_{\beta(Q+1)}^{NM}) + z_{Q+1}^{NM}, d_{\beta(Q+1)}^{NM}, t_{B1}^{NM}, d_{b1}^{NM}) \quad (24)$$

Define the block error probability as

$$\bar{P}_e = \Pr(\exists i \in \{2...Q+1\}, s.t. x_{Ai}^N \text{ is in error,}$$
$$\text{or } t_{D(i-1)}^N \text{ is in error, or } t_{D(Q+1)}^N \text{ is in error.}) \quad (25)$$

where $x_{Ai}^N$ is the part of $x_{Ai}^{NM}$ that is within one block. Similar notations are used for $t_{D(i-1)}^N$ and $t_{D(Q+1)}^N$. Given the signaling scheme presented in section IV-B and [17, Theorem 2], the probability of decoding error at each relay node goes to zero as $N \to \infty$. Let $P_e(i, k)$ be the probability of decoding error at relay node $i$ during phase $k$. Then $\bar{P}_e$ is related to $P_e(i, k)$ as $\bar{P}_e \leq 1 - \prod_{i,k}(1 - P_e(i, k))$, where the subscript in product includes the indices of all the relay node and the indices of the phases in this block.

For any given block length $Q$, we have $\lim_{N \to \infty} \bar{P}_e = 0$. Note that $\bar{P}_e$ is just a function of $N$ and $Q$. Because there are only finite number of relay nodes, this convergence is uniform over all relay nodes.

Let the equivocation under error free decoding be

$$\bar{H}_2 = \frac{1}{NM} H(W | (x_{A1}^{NM} \oplus d_{\alpha 1}^{NM}) + z_1^{NM}, d_{\alpha 1}^{NM}$$
$$(\bar{x}_{Ai}^{NM} \oplus d_{\alpha i}^{NM}) + (\bar{t}_{D(i-1)}^{NM} \oplus d_{\beta(i-1)}^{NM}) + z_i^{NM},$$
$$d_{\alpha i}^{NM}, d_{\beta(i-1)}^{NM}, i = 2...Q+1$$
$$(\bar{t}_{D(Q+1)}^{NM} \oplus d_{\beta(Q+1)}^{NM}) + z_{Q+1}^{NM}, d_{\beta(Q+1)}^{NM}, t_{B1}^{NM}, d_{b1}^{NM}) \quad (26)$$

where $\bar{x}_{Ai}^{NM}$ equals the value $x_{Ai}^{NM}$ takes with error free decoding. $\bar{t}_{D(i-1)}^{NM}$ and $\bar{t}_{D(Q+1)}^{NM}$ are defined in a similar fashion. Then we have the following lemma:

*Lemma 2:* For a given $Q$, $\bar{H}_2 + \varepsilon_2 \geq H_2 \geq \bar{H}_2 - \varepsilon_1$ where $\varepsilon_{1,2} \to 0$ as $N, M \to \infty$.

*Proof:* Let $c^j, \hat{c}^j$ denote the part of signals received by the relay node within the $j$th block. More specifically, they have the following form:

$$\hat{c}^j = \{(x_{Ai}^N(j) \oplus d_{\alpha i}^N(j)) +$$
$$(t_{D(i-1)}^N(j) \oplus d_{\beta(i-1)}^N(j)) + z_i^N(j), i = 2...Q+1\} \quad (27)$$
$$c^j = \{(\bar{x}_{Ai}^N(j) \oplus d_{\alpha i}^N(j)) +$$
$$(\bar{t}_{D(i-1)}^N(j) \oplus d_{\beta(i-1)}^N(j)) + z_i^N(j), i = 2...Q+1\} \quad (28)$$

In this notation, we exclude the first and the last batch of received signals. The first batch of received signals does not undergo any decoding operation. For the last batch of received signals we have the following notation:

$$\hat{f}^j = (t_{D(Q+1)}^N(j) \oplus d_{\beta(Q+1)}^N(j)) + z_{Q+1}^N(j) \quad (29)$$
$$f^j = (\bar{t}_{D(Q+1)}^N(j) \oplus d_{\beta(Q+1)}^N(j)) + z_{Q+1}^N(j) \quad (30)$$

The block index $(j)$ will be omitted in the following discussion for clarity.

We first prove that $c^j - \hat{c}^j$ is a discrete random variable with a finite support. According to the notation of (28), $c^j - \hat{c}^j$ has $Q$ components. Each component can be expressed as

$$(\bar{x}_{Ai}^N \oplus d_{\alpha i}^N) - (x_{Ai}^N \oplus d_{\alpha i}^N) +$$
$$(\bar{t}_{D(i-1)}^N \oplus d_{\beta(i-1)}^N) - (t_{D(i-1)}^N \oplus d_{\beta(i-1)}^N) \quad (31)$$

For the first line of (31) we have

$$(\bar{x}_{Ai}^N \oplus d_{\alpha i}^N) - (x_{Ai}^N \oplus d_{\alpha i}^N) \quad (32)$$
$$= \bar{x}_{Ai}^N + d_{\alpha i}^N + x_1^N - (x_{Ai}^N + d_{\alpha i}^N + x_2^N) \quad (33)$$
$$= \bar{x}_{Ai}^N - x_{Ai}^N + x_1^N - x_2^N \quad (34)$$

where $x_1^N, x_2^N$ belong to the coarse lattice $\Lambda_1$. Applying Theorem 1, we note that $x_1^N$ and $x_2^N$ each has at most $2^N$ possible solutions. $\bar{x}_{Ai}^N$ and $x_{Ai}^N$ each take $\|\mathcal{V}_1 \cap \Lambda\|$ possible values. Let $R = \frac{1}{N} \log_2 \|\mathcal{V}_1 \cap \Lambda\|$. Then (32) takes at most $2^{2N(R+1)}$ possible values. Similarly, we can prove that the second line of (31) has at most $2^{2N(R+1)}$ possible values as well. Therefore $c^j - \hat{c}^j$ takes at most $2^{4NQ(R+1)}$ possible values. Therefore $H(c^j - \hat{c}^j) \leq 4NQ(R + 1)$. Similarly, it

can be shown that $f - \hat{f}$ has at most $2N(R+1)$ solutions. This means that

$$H(c^j - \hat{c}^j, f^j - \hat{f}^j) \leq (4Q+2)N(R+1) \quad (35)$$

Let $c = \{c^j\}$, $\hat{c} = \{\hat{c}^j\}$, $f = \{f^j\}$ and $\hat{f} = \{\hat{f}^j\}$ $j = 1...M$. Let $b$ denote the remaining conditioning terms in $H_2$. Let $E^j$ denote the random variable $c^j \neq \hat{c}^j$ or $f^j \neq \hat{f}^j$. Then with probability $\bar{P}_e$ that $E^j = 1$. Otherwise $E^j = 0$. Let $W$ be the message transmitted over the $M$ blocks. Then we have

$$H(W|b, \hat{c}, \hat{f})$$
$$\geq H(W|b, c, \hat{c}, f, \hat{f}) \quad (36)$$
$$= H(W|b, c, f, c - \hat{c}, f - \hat{f}) \quad (37)$$
$$= H(W|b, c, f) + H(c - \hat{c}, f - \hat{f}|W, b, c, f)$$
$$\quad - H(c - \hat{c}, f - \hat{f}|b, c, f) \quad (38)$$
$$\geq H(W|b, c, f) - H(c - \hat{c}, f - \hat{f}) \quad (39)$$
$$\geq H(W|b, c, f) - \sum_{j=1}^{M} H(c^j - \hat{c}^j, f^j - \hat{f}^j) \quad (40)$$
$$= H(W|b, c, f) - \sum_{j=1}^{M} H(c^j - \hat{c}^j, f^j - \hat{f}^j, E^j) \quad (41)$$
$$\geq H(W|b, c, f) - \sum_{j=1}^{M} H(E^j)$$
$$\quad - \sum_{j=1}^{M} \Pr(E^j = 1) H(c^j - \hat{c}^j, f^j - \hat{f}^j) \quad (42)$$
$$\geq H(W|b, c, f) - M - M\bar{P}_e(4Q+2)N(R+1) \quad (43)$$

By dividing $NM$ on both sides and letting $N, M \to \infty$, and $\varepsilon_1 = 1/N + \bar{P}_e(4Q+2)(R+1)$ we get $H_2 \geq \bar{H}_2 - \varepsilon_1$. Similarly we can prove $\bar{H}_2 \geq H_2 - \varepsilon_2$. ∎

*Remark 3:* Lemma 2 says that if a particular equivocation value is achievable with regard to one relay node, when all the other relay nodes do error free decoding, then the same equivocation value is achievable when other relay nodes do decode and forward which is only error free in asymptotic sense.

*Lemma 3:* $\bar{H}_2$ is the same for all relay nodes.

*Proof:* Lemma follows because relay nodes receive statistically equivalent signals if there are no decoding errors. For the $k$th relay node, as shown by the edge labels in Figure 3, the condition term of $\bar{H}_2$ in (26) is related to $t_j^{NM}$ as follows:

$$x_{A1}^{NM} = J_{k-2}^{NM} \quad (44)$$
$$\bar{x}_{A2}^{NM} = t_0^{NM} \oplus J_{k-1}^{NM} \quad (45)$$
$$\bar{x}_{A3}^{NM} = t_0^{NM} \oplus t_1^{NM} \oplus J_k^{NM} \quad (46)$$
$$...$$
$$\bar{x}_{A(Q+1)}^{NM} = t_0^{NM} \oplus t_1^{NM} \oplus ... \oplus t_{Q-1}^{NM} \oplus J_{K+Q-2}^{NM} \quad (47)$$
$$\bar{t}_{D1}^{NM} = J_k^{NM} \quad (48)$$
$$\bar{t}_{D2}^{NM} = t_0^{NM} \oplus J_{k+1}^{NM} \quad (49)$$
$$\bar{t}_{D3}^{NM} = t_0^{NM} \oplus t_1^{NM} \oplus J_{k+2}^{NM} \quad (50)$$
$$...$$
$$\bar{t}_{D(Q+1)}^{NM} = t_0^{NM} \oplus t_1^{NM} \oplus ... t_{Q-1}^{NM} \oplus J_{k+Q}^{NM} \quad (51)$$
$$t_{B1}^{NM} = J_{k-1}^{NM} \quad (52)$$

Given the lattice points transmitted by the source $t_j^{NM}$, the joint distribution of the side information for any relay node is the same. Hence we have the lemma. ∎

With these preparation, we are now ready to present the following achievable rate.

*Theorem 4:* For any $\varepsilon > 0$, a secrecy rate of at least $0.5(C(2\bar{P} - 0.5) - 1) - \varepsilon$ bits per channel use is achievable regardless of the number of hops.

*Proof:* According to Lemma 3, it suffices to design the coding scheme based on one relay node. We focus on one block of channel uses as shown in Figure 3. Let $V(j)$ to denote all the side information available to the relay node within the $j$th block. We start by lower bounding $H(t_0^{NQ}|V(j))$ under ideal error free decoding, where $t_0^{NQ}$ are the lattice points picked by the encoder at the source node as described in Section IV-B within this block. $H(t_0^{NQ}|V(j))$ equals

$$H(t_0^{NQ}|(\bar{x}_{Ai}^N \oplus d_{\alpha i}^N) + (\bar{t}_{D(i-1)}^N \oplus d_{\beta(i-1)}^N) + z_i^N,$$
$$d_{\alpha i}^N, d_{\beta(i-1)}^N, i = 2...Q+1, t_{B1}^N, d_{b1}^N) \quad (53)$$

Comparing (53) with the condition terms in (26), we see that we have removed the first batch and the last batch of received signals during a block from the condition terms because they are independent from everything else. The last batch of received signals contains the lattice point of the most recent jamming signal observable by the relay node. Its independence follows from Lemma 1.

We then assume that the eavesdropper residing at the relay node knows the channel noise. This means (53) can be lower bounded by:

$$H(t_0^{NQ}|(\bar{x}_{Ai}^N \oplus d_{\alpha i}^N) + (\bar{t}_{D(i-1)}^N \oplus d_{\beta(i-1)}^N),$$
$$d_{\alpha i}^N, d_{\beta(i-1)}^N, i = 2...Q+1, t_{B1}^N, d_{b1}^N) \quad (54)$$

Next, we invoke Theorem 1. Equation (54) can be lower bounded by:

$$H(t_0^{NQ}|\bar{x}_{Ai}^N \oplus d_{\alpha i}^N \oplus \bar{t}_{D(i-1)}^N \oplus d_{\beta(i-1)}^N, T_i,$$
$$d_{\alpha i}^N, d_{\beta(i-1)}^N, i = 2...Q+1, t_{B1}^N, d_{b1}^N) \quad (55)$$

where $T_i$ can be represented with $N$ bits. Using the similar argument as in (9)-(13), (55) is lower bounded by:

$$H(t_0^{NQ}|\bar{x}_{Ai}^N \oplus d_{\alpha i}^N \oplus \bar{t}_{D(i-1)}^N \oplus d_{\beta(i-1)}^N,$$
$$d_{\alpha i}^N, d_{\beta(i-1), i=2...Q+1}^N, t_{B1}^N, d_{b1}^N) - H(T_{i, i=2...Q+1}) \quad (56)$$
$$= H(t_0^{NQ}|\bar{x}_{Ai}^N \oplus \bar{t}_{D(i-1), i=2...Q+1}^N, t_{B1}^N) - H(T_{i, i=2...Q+1}) \quad (57)$$

It turns out that in the first term in (57), the conditional variables are all independent from $t_0^{NQ}$. This is because $\bar{t}_{D(i-1)}^N$ contains $J_{i-2+k}^N$, which is a new lattice point not contained in previous $\bar{t}^N_{D(j-1)}$ or $\bar{x}^N_{Aj}$ $j < i$. The new lattice point is uniformly distributed over $\mathcal{V}_1 \cap \Lambda$. Therefore, from Lemma 1, $\bar{x}^N_{Ai} \oplus \bar{t}^N_{D(i-1)}$ is independent from $t^{NQ}_0$. Therefore (57) equals

$$H(t^{NQ}_0) - H(T_{i,i=2...Q+1}) \qquad (58)$$

Define

$$c = \frac{1}{NQ} I(t^{NQ}_0; V(j)) \qquad (59)$$

Then from (58), we have $c \in (0, 1)$.

To achieve perfect secrecy, a similar argument of coding across different blocks as the one in Section III can be used. A codebook with rate $R$ and size $2^{\lfloor MNQR \rfloor}$ that spans over $M$ blocks is constructed as follows: Each codeword is a length $MQ$ sequence. Each component of the sequence is an $N$-dimensional lattice point sampled in an i.i.d fashion from the uniform distribution over $\mathcal{V}_1 \cap \Lambda$. The codebook is then randomly binned into several bins. Each bin contains $2^{\lfloor MNQc \rfloor}$ codewords, with $c$ given by (59). Denote the codebook with $\mathcal{C}$.

The transmitted codeword is determined as follows: Consider a message set $\{W\}$, whose size equals the number of the bins. The message is mapped to the bins in a one-to-one fashion. The actual transmitted codeword is then selected from the bin according to a uniform distribution. Let this codeword be $u^{MNQ}$. Let $V = \{V(j), j = 1...M\}$. Then we have:

$$H(W|V, \mathcal{C}) \qquad (60)$$
$$= H(W|u^{MNQ}, V, \mathcal{C}) + H(u^{MNQ}|V, \mathcal{C})$$
$$\quad - H(u^{MNQ}|W, V, \mathcal{C}) \qquad (61)$$
$$\geq H(u^{MNQ}|V, \mathcal{C}) - MNQ\varepsilon \qquad (62)$$
$$= H(u^{MNQ}|\mathcal{C}) - I(u^{MNQ}; V|\mathcal{C}) - MNQ\varepsilon \qquad (63)$$
$$\geq H(u^{MNQ}|\mathcal{C}) - \sum_{j=1}^{M} I(u^{MNQ}(j); V(j)) - MNQ\varepsilon \qquad (64)$$
$$= H(u^{MNQ}|\mathcal{C}) - MNQc - MNQ\varepsilon \qquad (65)$$

(62) follows from Fano's inequality and the size of the bin is picked according to the rate of information leaked to the eavesdropper under the same input distribution used to sample the codebook. (64) follows from $\mathcal{C} \to u^{MNQ} \to V$ being a Markov chain. Divide (60) and (65) by $MNQ$ and let $M \to \infty$, we have $\varepsilon \to 0$ and $\lim_{M \to \infty} \frac{1}{MNQ} H(W|V, \mathcal{C}) = \lim_{M \to \infty} \frac{1}{MNQ} H(W)$. Therefore a secrecy rate of $R - c$ bits per channel use is achieved. According to [10], $R$ can be arbitrarily close to $C(P - 0.5)$ by making $N \to \infty$, where $P$ is the average power per channel use spent to transmit a lattice point. For a given node, during $2Q + 3$ phases, it is active in $Q + 1$ phases. Since $c \in [0, 1]$, a secrecy rate of $\frac{Q+1}{2Q+3}(C(\frac{2Q+3}{Q+1}\bar{P} - 0.5) - 1)$ is then achievable by letting $M \to \infty$. Taking the limit $Q \to \infty$, we have the theorem. ∎

## V. CONCLUSION

Lattice codes were shown recently as a useful technique to prove information theoretic results. In this work, we showed that lattice codes are also useful to prove secrecy results. This was done by showing that the equivocation rate could be bounded if the shaping set and the "fine" lattice forms a nested lattice structure. With this new tool, we computed the secrecy rate for two models: (1) a wiretap channel with a cooperative jammer, (2) a multi-hop line network with untrusted relays. For the second model, we have shown that a coding scheme can be designed to support a non-vanishing secrecy rate regardless of the number of hops.